\documentclass[pre,showpacs,preprint]{revtex4}
\usepackage{amsmath}

\usepackage{graphicx}
\usepackage{dcolumn}
\usepackage{bm}

\begin{document}

\title{Hydrodynamic Character of the  Non-equipartition of Kinetic Energy in Binary Granular Gases}
\author{J. Javier Brey and M.J. Ruiz-Montero}
\affiliation{F\'{\i}sica Te\'{o}rica, Universidad de Sevilla,
Apartado de Correos 1065, E-41080, Sevilla, Spain}
\date{\today }

\begin{abstract}
The influence of the heating mechanism on the kinetic energy densities of the components of a vibrated granular mixture is investigated. Collisions of the particles with the vibrating wall are inelastic and characterized by two coefficients of normal restitution, one for each of the two species. By means
of molecular dynamics simulations, it is shown that the non-equipartition of kinetic energy is not affected by the
differential mechanism of energy injection, aside the usual boundary layer around the wall. The macroscopic state
of the mixture in the bulk is defined by intensive variables that do not include the partial granular temperatures
of the components.
\end{abstract}

\pacs{45.70.-n,47.70.Nd,51.10.+y}

\maketitle

A fundamental issue for non-equilibrium physics is the identification of the variables needed to characterize a given macroscopic state \cite{Zw01}. Granular gases are a widely studied example of out of equilibrium systems \cite{ByP04}. Due to the
inelasticity of collisions, they do not have an equilibrium state and, in order to keep the granular medium fluidized, external energy must be continuously supplied in some way. An extension of the usual hydrodynamic equations for
molecular gases has proven to provide an accurate description of granular gases in many different contexts \cite{Go03}. This description involves the granular temperature, defined from the mean kinetic energy of the particles, just by analogy with the definition of temperature in kinetic theory of molecular fluids.

For granular mixtures, one of the consequences of inelasticity is that the energy equipartition required in equilibrium systems is not verified. The granular temperatures of the components of a mixture, defined from the average kinetic energy of each species, are different. This feature was pointed out many years ago \cite{JyM87}, and a systematic study of it has started in the last years. Once established that the partial temperatures of the several components of the mixture are different, some important conceptual questions arise: should they be incorporated into the set of intensive variables needed to identify a given macroscopic state of a granular mixture, replacing the single temperature field employed in the description of one-component systems? If the answer is negative, are the partial temperatures determined by the usual set of hydrodynamic fields (including the temperature of the mixture) and the properties of the two types of particles? It is important to realize that for ordinary mixtures, the partial temperatures of the components
are not included in the hydrodynamic description. Moreover, kinetic theory then implies that the temperature fields of all the components are the same, i.e. equipartition applies locally \cite{FyK72}.

In the theoretical study of granular mixtures, it is usually assumed that the temperature fields of each of the especies are not needed to describe the system \cite{GyD99}. In a kinetic theory or statistical mechanics description, this implies that the temperatures of the components can be determined from the other macroscopic parameters, as indicated above, since the (normal) distribution function of the system is assumed to be determined by the hydrodynamic fields, so all the properties of the system turn out to be functionals of them \cite{Zw01}.

Recently, Wang and Menon \cite{WyM08} reported the results of some event-driven simulations of a granular mixture
and reached the conclusion that the heating mechanism affects nonequipartition of energy, even in the bulk of the system. This would imply that the details of the driving used to inject energy into the system can not be ignored
in describing mixtures of inelastic gases even far away from the energy source. If this were actually the case, the generalized hydrodynamic-like equations for granular gases should include the partial temperatures of the components, and the existing derivations of them should be deeply revised. In this paper, the influence of the differential boundary heating on the violation of equipartition is reconsidered by means of molecular dynamics simulations. It is found that the details of the driving mechanism do not affect the relations between intensive quantities aside from a boundary layer next to the energizing wall.

In the event-driven simulations whose results will be described below, the system is composed of two kinds of smooth inelastic hard disks, labeled $1$ and $2$. Both types of disks have the same diameter $\sigma$, but different masses, $m_{1}$ and $m_{2}$. The number of particles of each species is $N_{1}$ and $N_{2}$, respectively, being the total number of particles $N=N_{1}+N_{2} = 420$, in all cases.
There is an external  gravitational field acting on the system, so that each particle $i$ is subjected to a force
$-m_{i} g_{0} \widehat{\bm e}_{z}$, where $g_{0}$ is a positive constant and $\widehat{\bm e}_{z}$
is the unit vector in the direction of the positive $z$ axis. The inelasticity of collisions is
modeled by constant coefficients of normal restitution. There are three of them: $\alpha_{11}$, $\alpha_{22}$, and
$\alpha_{12}=\alpha_{21}$, where $\alpha_{ij}$ refers to collisions between a particle of species $i$ and a particle of  species $j$. They are defined in the interval $0 < \alpha_{ij} \leq 1$. For the simulations reported here, $\alpha_{12}$ has been chosen as given by $\alpha_{12}=(\alpha_{11}+\alpha_{22})/2$. The system is open on
the top and its width is $W=70 \sigma$. The latter value has been chosen small enough so that the steady state with gradients only in the vertical ($z$) direction is stable, avoiding the development of transversal instabilities
\cite{SyK01,LMyS02,BRMyG02}.  To eliminate undesired boundary effects induced by the lateral walls, periodic boundary conditions in the horizontal ($x$) direction have been employed.

To keep the system fluidized, energy is continuously supplied through the wall located at the bottom ($z=0$) of
the system. The specific way in which this wall is modeled has been chosen from a compromise between several
considerations. The physical issue of interest, unequally heating of the two species, should be isolated, avoiding additional effects on the properties of interest induced, for instance, by the oscillations of the wall. Also, the heating should be, as much as possible, the (idealized) model of some experimental process. In this context, ``thermal'' walls in which the velocity of a reflected particle is not related with its incident velocity, are not at all neutral
for the ``adjustment''  of the dispersion of the velocities of the two species, since they destroy all the velocity correlations by definition. In addition, it does not seem clear to which type of experimental boundary condition for granular fluids they correspond to, if any. Thus a vibrating wall with a sawtooth velocity profile is chosen, so that all the particles
colliding with the wall find it with the same upwards velocity $v_{W}$. Moreover, the wall is supposed to move with an amplitude much smaller than the mean free path of the particles in its proximity. Consequently, an accurate description is obtained by considering it as fixed at $z=0$ \cite{McyB97,McyL98}. Therefore, the dynamics of this wall does not induce any
additional space or time dependence. The mechanism for the differential heating must be now introduced.
It will be attached to the difference in the coefficient of normal restitution, $\alpha_{W1}$ and $\alpha_{W2}$, for the collisions of the two types of particles with the vibrating wall. When a particle of species $i$ having a velocity ${\bm v}$ collides with the wall, its component $v_{x}$ remains unchanged, while the $z$ component is
instantaneously modified to
\begin{equation}
\label{2.1} v_{z}^{\prime}=v_{z}- (1+\alpha_{Wi})(v_{z}-v_{W}).
\end{equation}
Note that a collision of this kind is only possible if $v_{z} <0$, since the wall is treated as located at $z=0$.
From Eq. (\ref{2.1}), it is obtained that
\begin{equation}
\label{2.2}
v_{z}^{\prime 2}- v_{z}^{2} = \left(1+\alpha_{Wi} \right) (v_{W}+|v_{z}|) \left[ (1+\alpha_{Wi})v_{W} -(1-\alpha_{W}) |v_{z}| \right].
\end{equation}
Therefore, when a particle collides with the vibrating wall, its kinetic energy decreases as a consequence of the collision if
\begin{equation}
\label{2.3}
|v_{z}| > v_{c} \equiv \frac{1+\alpha_{Wi}}{1-\alpha_{Wi}}\, v_{W}.
\end{equation}
When the average velocity of the particles of one of the species approaching the vibrating wall is larger than $v_{c}$, this species loses energy in the collisions with the wall, on the average.
Of course, the net balance of energy flux through the vibrating wall when considering both species must be positive since, in the steady state, energy
must be continuously supplied by the wall to compensate for the dissipation in collisions. Note that this effect, cooling of one of the species by the vibrating wall, is not possible
when the particles collide with the wall in an elastic way, as $v_{c}$ diverges in this case. On the other hand,
it can occur when thermal walls are considered, as made for instance in \cite{WyM08}.

\begin{figure}
\includegraphics[scale=0.5,angle=0]{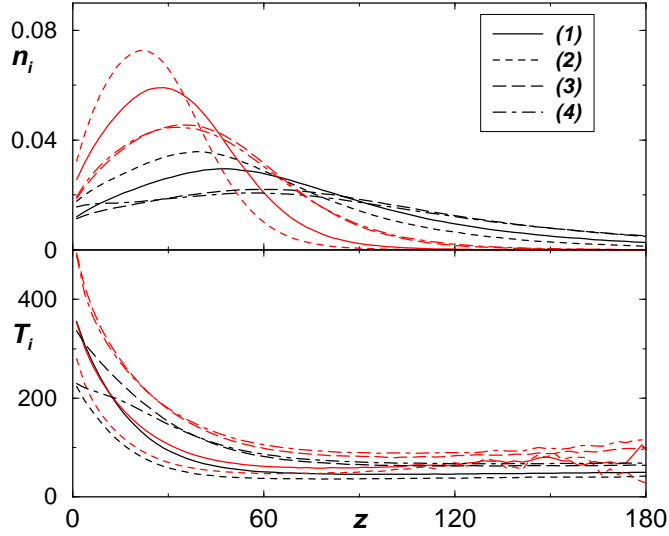}
\caption{(color online) Number density, $n_{i}$, measured in units of $ \sigma^{-2}$, and granular temperature, $T_{i}$, measured in  units
of $m_{1} \sigma g_{0}$, profiles  for each of the the two species of particles, along the vertical direction. Height is measured in units of $\sigma$. The masses of the two kinds of particles are related by $m_{2}= 5 m_{1}$. There are $210$ particles of each species and the width of the system is $W= 70 \sigma$.  Four levels of differential heating at the vibrating wall, labeled from $1$ to $4$, are shown; they are specified in the main text. The density profiles for species $2$ (red online) exhibit higher maxima than those of species $1$ (black online). Each temperature profile for species $2$ (red online) is always above the profile for specie $1$ (black online) corresponding to  the same driving. }
\label{fig1}
\end{figure}

Four differential drivings will be considered in the following: (1) $\alpha_{W1}=0.999$ and $\alpha_{W2}=0.5$,
(2) $\alpha_{W1}=0.995$ and $\alpha_{W2}=0.9$, (3) $\alpha_{W1}=0.9$ and  $\alpha_{W2}=0.995$, and (4) $\alpha_{W1}=0.5$ and $\alpha_{W2}=0.999$. In Fig. \ref{fig1} the density, $n_{i}(z)$, and temperature, $T_{i}(z)$, profiles for each of the two species in the steady state are shown for each of the above drivings. In all the simulations reported in the figure, $\alpha_{11}= \alpha_{22}= 0.93$, $m_{2}=5m_{1}$, and $N_{1}=N_{2}$. These are the same values as considered in ref. \cite{WyM08}. The velocity of the vibrating wall, $v_{W}$, has been chosen in each case such that the system remains in the dilute regime. The specific values used for each of the above drivings are: $v_{W} / \sqrt{g_{0} \sigma}=7 $ for
driving $(1)$, $4$ for driving $(2)$, $5$ for driving $(3)$, and $6$ for driving $(4)$. To illustrate how the mechanism of energy injection  actually affects to each of the two species, the average variation of the kinetic energy of the particles when colliding with the vibrating wall in the steady situation, $<\Delta e_{i}>$,  has been computed for the two kind of particles. The value of the ratio $m_{1}<\Delta e_{2}>/ m_{2}<\Delta e_{1}>$ for each of the four drivings specified above is: $0.19$ for driving $(1)$, $0.47$ for driving $(2)$, $0.85$ for driving $(3)$, and $-6.20$ for driving $(4)$. The disparity in the above values, clearly shows that the input of energy per unit of mass
for each of the components depends very strongly on the particular differential driving being used. The negative sign for driving (4) is due to the partial cooling of species $1$ upon colliding with the vibrating wall, that corresponds to the possible scenario pointed out above.

In Fig. 1, it is seen that the profiles of the hydrodynamic fields are different for the different drivings, everywhere in the system.
Actually, the same happens if the density ratio, $n_{2}(z)/n_{1}(z)$, or the temperature ratio, $\gamma (z) \equiv T_{2}(z)/T_{1}(z)$, are plotted. This is not at all surprising, since the hydrodynamic profiles and also the density and temperature ratios are expected to depend on the boundary conditions. In monodisperse systems with an elastic vibrating wall, it has been shown that  $v_{W}^{2}$ plays the role of an scaling factor for the amplitude of the hydrodynamic fields of the gas in the steady state \cite{BRyM01}. Mathematically, this property follows by realizing that the energy  injected by the wall into the system is $\Phi_{W}= Nmg_{0}v_{W}/W$, and it is a consequence of the inelastic Navier-Stokes hydrodynamic description. From this point of view, the verification of the scaling in a region can be considered as a signature of the bulk of the system.

For the two-component system with differential heating we are considering, it can be expected on physical grounds that the relevant quantity characterizing the vibrating boundary at a macroscopic level is again the injected energy flux, and that in the low density limit it scales the hydrodynamic fields in the same way as  $v_{W}$ does in the one-component case. Then, dimensionless temperature and density fields are defined by
\begin{equation}
\label{2.4}
n^{*}_{i}= \frac{n_{i} \Phi_{W}^{2} \sigma^3}{m_{1}^2 g_{0}^{3}}\, ,
\end{equation}
and
\begin{equation}
\label{2.5}
T^{*}_{i}= \frac{T_{i} m_{1} g_{0}^{2}}{\sigma^{2} \Phi_{W}^{2}}\, ,
\end{equation}
respectively.  Consistently, an scaled coordinate $z^{*}$ is introduced by
\begin{equation}
\label{2.6} z^{*}= \frac{z m_{1}^{2} g_{0}^{3}}{\sigma^{2} \Phi_{W}^{2}}\, .
\end{equation}

\begin{figure}
\includegraphics[scale=0.5,angle=0]{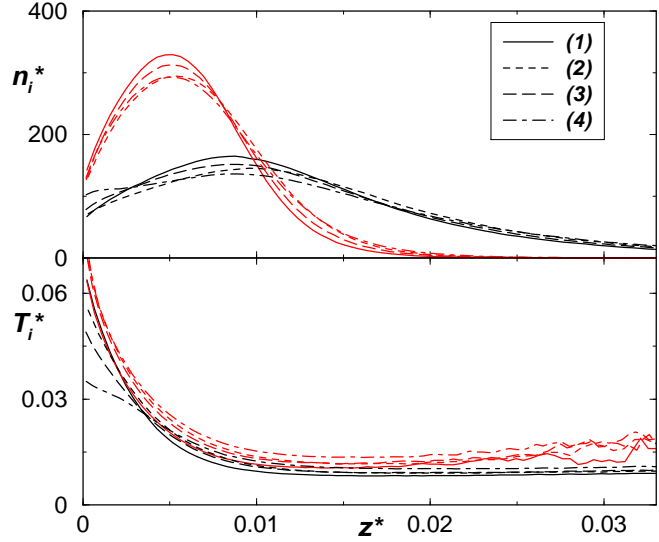}
\caption{(color online) Dimensionless number density, $n^{*}_{i}$, and granular temperature, $T^{*}_{i} $ profiles,
defined in Eqs. (\protect{\ref{2.4}}) and (\protect{\ref{2.5}}),  for the two species of particles along the vertical direction, in the dimensionless length scale introduced by Eq. (\protect{\ref{2.6}}). The system and the several differential drivings are the same as in Fig.\ \protect{\ref{fig1}}.}
\label{fig2}
\end{figure}

The scaled density and temperature profiles in the new length scale for the same systems as in Fig. \ref{fig1} are plotted
in Fig. \ref{fig2}. The measured values of $\Phi_{W}$ for each of the four drivings, measured in units of $m g_{0}^{3/2} \sigma^{-1/2}$, are: $74.62$ for driving $(1)$, $63.67$ for driving $(2)$, $82.83$ for driving $(3)$, and $81.00$ for driving $(4)$. Although the curves do not overlap perfectly, it seems fair to conclude that there
exist well defined scaled profiles for each of the components, aside from a boundary region next to the
vibrating wall. Actually, the observed discrepancies can be due to finite density effects, similar to those found
in \cite{ByR09}, since the maximum value of total number density reaches values $n  \equiv n_{1}+n_{2} \sim 0.1$, which are not very low. In this context, it is worth to insist that the scaling for one-dimensional systems only holds in the dilute
limit.

\begin{figure}
\includegraphics[scale=0.5,angle=0]{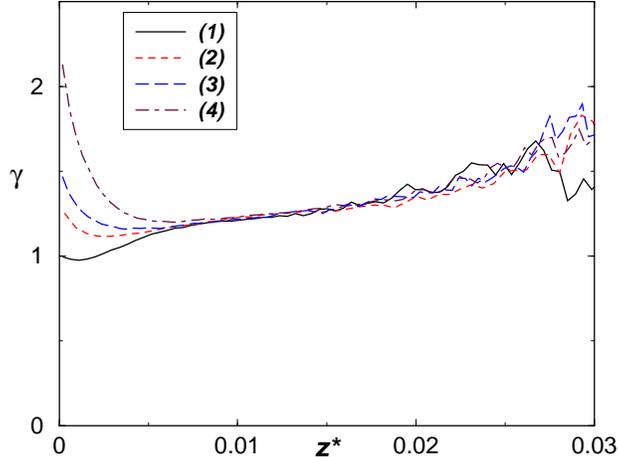}
\caption{(color online) Temperature ratio profile, $\gamma$, for the
two species of particles along the vertical direction. The system and the differential heatings are the same as
in Fig.\ \protect{\ref{fig1}}. The vertical $z^{*}$ coordinated is measured in the dimensionless units defined
in Eq. (\protect{\ref{2.6}}), which involve both the parameters defining the system and the particular heating being used.}
\label{fig3}
\end{figure}

Now that the appropriate scaling to compare the hydrodynamic profiles has been identified, it can be used to check
the hydrodynamic nature of $\gamma$, in the sense that, in the bulk of the system, it is  determined by local properties of the hydrodynamic fields of the mixture. In Fig. \ref{fig3}, the profile of $\gamma$ is plotted as a function of the dimensionless scale $z^{*}$. It is observed that the curves corresponding to the four differential heating mechanisms collapse in a wide region of the system, actually in most of it since the fluctuations for large values of $z^{*}$ are due to the small number of particles present there. This collapse is a very strong indication of the hydrodynamic character of $\gamma$, and it shows that the departure from equipartition is determined by the local properties of the mixture, being unnecessary to introduce
independent temperature fields to describe the behavior of each of the two species. It is important to
realize the relevance of the scaling when investigating the bulk properties of $\gamma$. Although trivially it does not affect the ratio, the value of $z^{*}$ following from a given $z$ depends, not only on the parameters defining the system itself, but also on the differential heating being used, since both determine the amount of power injected through the vibrating wall in the steady state. It can be wondered why the collapse of the $\gamma$ profiles is clearer
than the collapse of the partial temperatures showed in Fig. \ref{fig2}. A probable reason is that finite density effects are more relevant for each of the component temperatures than for its ratio. The same kind of collapse was
obtained for other systems with mass ratio between $0.5$ and $5$, and with values of the coefficients of normal restitution up to roughly $0.8$. For smaller values of these coefficients, the one-dimensional hydrodynamical profiles become unstable and a more involved analysis is needed.

In summary, the simulation results reported here show that when energy is supplied to a granular mixture through a vibrating wall, the details of the collision mechanism for each of the species do not affect the extent of non-equipartition, beyond the usual boundary layer. This means that a macroscopic description does not require to introduce the partial (granular) temperatures of the components. We expect a similar conclusion to hold for other kind of differential boundaries and also for sheared granular fluids. Of course, a different question is the theoretical prediction of the departure from equipartition, measured for instance by the partial  temperature ratio. The theoretical studies carried out up to now \cite{GyD99,ByT02,ByP02} refer to homogeneous systems, with the exception of the dilute tracer limit in a vibrated system \cite{BRyM05,BRyM06}, although the results have been compared to experimental data obtained in spatially inhomogeneous vibro-fluidized systems, and a relative qualitative agreement has been found \cite{Lo99,FyM02,WyP02}.

This research was supported by the Ministerio de Educaci\'{o}n y
Ciencia (Spain) through Grant No. FIS2008-01339 (partially financed
by FEDER funds).

\end{document}